\documentclass[12pt]{article}

\newcommand{\Tr}{\mathop{\rm Tr}\nolimits}

\title{Spin calculations in arbitrary and diagonal spin bases%
\thanks{Presented at the VIth International School-seminar
{\it``Actual Problems of High Energy Physics''}. Gomel, Belarus,
August 7-16, 2001}}

\author{Sergey~M.~Sikach
\\ \sl B.I.Stepanov Institute of Physics of \\ \sl National
Academy of Sciences of Belarus  \and \sl F.Skorina Av., 68, Minsk,
220072, Republic of Belarus
 \and \rm Tel: (375-17) 284-04-29  \and \rm  Fax: (375-17) 284-08-79
  \and \rm e-mail: sikach@dragon.bas-net.by}

\date{}

\begin{document}
\maketitle

\begin{abstract}
We consider the  spin calculations in arbitrary and diagonal spin
bases
\end{abstract}

%%%%%%%%%%%%%%%%%%%%%%%%%%%%%%%%%%%%%%%%%%%%%%%%%%%%%%%%%%%%%%%%%%%%%%

\section {Introduction}%1

 At the present time, the conventional method of
direct calculation of the squared modulus of amplitudes is
practically not used in calculating of reactions. This is due to
the fact that even an insignificant increase in the number of
particles participating in a reaction as well as the account of
their polarizations sharply complicate calculations. The idea of
direct calculation of amplitudes followed by their quadrating is
fairly obvious. However, such an approach entails a number of
specific problems. Therefore, in the last few decades publications
devoted to the development of particular aspect of this method
have been appearing in the press (see references in reviews
\cite{r.1}, \cite{r.2}).

These works can conventionally be classified according to the
following directions: a covariant or an incovariant approach is
used; calculations are carried out in an arbitrary or a particular
spin basis. Among the first works devoted to this topic are the
works of F.I.~Fedorov and his scholars. For details see
\cite{r.3}. They are devoted primarily to covariant calculation of
amplitudes in an arbitrary spin basis. For instance, in \cite{r.4}
a method that permits using in calculating amplitudes the main
``trick'' of the quadrating procedure was proposed. Namely,
reduction of calculations to the computation of traces from
$\gamma$-matrix operators.

Let us illustrate this approach by an example of a fermion
``sandwich'' which can be given in the following form ($Q$ --
arbitrary operator):
\begin {equation}
\displaystyle
\bar{u}^{{\sigma}'} ( p' , s' ) Q u^{\sigma}( p , s ) = \Tr Q
u^{\sigma}( p , s ) \bar{u}^{{\sigma}'} ( p' , s' ) \; .
\label{e.1}
\end {equation}

The main difficulties of this approach are the obtaining of an
explicit form of the transition operator
\begin {equation}
\displaystyle
u^{\sigma} ( p , s ) \bar{u}^{{\sigma}'} ( p' , s' ) \; ,
\label{e.2}
\end {equation}
as well as the solution of such problems as the calculation of the
phase factors, the removal of singularities, the calculation of
the exchange diagrams, and the account of the specificity of the
spin configuration $\sigma' = - \sigma$, and so on. Some of these
difficulties can be overcome by fixing a certain manner the spin
basis. Transition to another basis is realized with the use of the
Wigner $D$-functions. However, this apparatus is exceedingly
cumbersome and incovariant.

The present paper shows how, using amplitudes (\ref{e.1}) in the
basis in which they have the simplest form, to obtain expressions
for the squared modulus of an amplitude in an arbitrary spin
basis. The paper considers the following questions. In Section
{\bf 2}, one-particle states are considered. It is shown that
consistent use of tetrads formalism permits operating with a
complete set of bispinors as well as reducing the operations of
matrix operators to tensor ones.

In Section {\bf 3}, matrices that make it possible, using the
amplitudes calculated in one spin basis, to calculate the squared
modulus of amplitudes in another spin basis.

Section {\bf 4} considers the introduced into \cite{r.5} the
diagonal spin basis (DSB). In the DSB, the bispinors of the
initial and final states have a common set of spin operators and
interrelated in the simplest way. As a result, the central problem
of this approach -- the obtaining of a simple and compact
solution. The diagonal amplitudes have a clear physical
interpretation and are the best ``building bricks'' for
calculation of squared modulus of amplitudes in arbitrary spin
basis.

In Section {\bf 5}, some examples of calculation in DSB are
considered.

%%%%%%%%%%%%%%%%%%%%%%%%%%%%%%%%%%%%%%%%%%%%%%%%%%%%%%%%%%%%%%%%%%%%%%

\section {One-particle states. Tetrads}%2

The one-particle state is defined by 4-vectors, momentum $p$ and
the axis of spin projections $s$. And $p s = 0$. This condition is
fulfilled if $s$ is determined in terms of the arbitrary vector
$q$
\begin {equation}
\displaystyle
s = {(p q) p - m^2 q \over m \sqrt{(p q)^2 - m^2 q^2  } }  \, .
\label{e.3}
\end {equation}
Define
\begin {equation}
\displaystyle
s_0 = v = {p  \over m }  \, , \; \; \; s_3 = s \, ,
\label{e.4}
\end {equation}
then $s_0^2 = - s_3^2 = 1$, $s_0 s_3 = 0$. Any vector having the
form $\alpha s_0 + \beta s_3$ belongs to the 2-plane $(s_0, s_3)$
in Minkowski space. The orthogonal 2-plane can be determined with
the aid of two tensors
\begin {equation}
\displaystyle
g^{\mu \nu}_{\parallel} = v^{\mu} v^{\nu} - s^{\mu} s^{\nu}  \, ,
\; \; \; \tilde{\varepsilon}^{\mu \nu}_{\parallel} =
{\varepsilon}^{\mu \nu \rho \sigma} v_{\rho} s_{\sigma} \, ,
\label{e.5}
\end {equation}
where $\varepsilon_{0 1 2 3} = 1$. If the arbitrary vector $r$
lies incompletelly in the 2-plane $(s_0, s_3)$, then the vectors
($g$ is Minkowski tensor)
\begin {equation}
\displaystyle
s_1^{\mu} =( g^{\mu \nu}_{\parallel} - g^{\mu \nu}) {r_{\nu} \over
r_{{}_\bot} } \, ,  \; \; \; s_2^{\mu} = -\tilde{\varepsilon}^{\mu
\nu}_{\parallel} {r_{\nu} \over r_{{}_\bot} } \, ;  \; \; \;
r_{{}_{ \bot} } = \sqrt{ r_{\mu} ( g^{\mu \nu}_{\parallel} -
g^{\mu \nu}) r_{\nu} }
\label{e.6}
\end {equation}
satisfy the conditions $s_1^2 =  s_2^2 = -1$, $s_1 s_2 = 0$ and
both vectors are orthogonal to $s_0$ and $s_3$. Thus, the tetrad
$s_0, s_3, s_1, s_2$ form orthonormalised basis in the Minkowski
space. In all spin reactions, vectors $s_1$ and $s_2$ only occur
in combination $s_1 \pm i s_2$. From (\ref{e.6}) follows
\begin {equation}
\displaystyle
s_1 + i \sigma s_2 = T_{\sigma} {r \over r_{{}_\bot} } \, , \; \;
\; T_{\sigma} = g_{\parallel} - g - i \sigma
\tilde{\varepsilon}_{\parallel} \, .
\label{e.7}
\end {equation}
It is clear that the tensor $T_{\sigma}$ plays the special role in
spin calculations. Besides the apparent properties
$T_{\sigma}^{\ast} = T_{-\sigma}$,  $T_{\sigma}^{+} = T_{\sigma}$,
$T_{\sigma}^2 = - 2 T_{\sigma}$, the relation
\begin {equation}
\displaystyle
 T_{\sigma}^{\mu \nu} T_{\sigma}^{\alpha \beta}  =
 T_{\sigma}^{\mu \beta} T_{\sigma}^{\alpha \nu}
\label{e.8}
\end {equation}
holds for it. From this relation follows
\begin {equation}
\displaystyle
 T_{\sigma}^{\mu \nu} r_{\nu}  r_{\alpha} T_{\sigma}^{\alpha \beta}  =
 r_{{}_\bot}^2  T_{\sigma}^{\mu \beta} \, .
\label{e.9}
\end {equation}
If in (\ref{e.7}) we replace the vector $r$ by $r'$, then from
(\ref{e.7}) and (\ref{e.9}) we obtain
\begin {equation}
\displaystyle
{s_1}' + i \sigma {s_2}' = e^{i \sigma \varphi} (s_1 + i \sigma
s_2) \, , \; \; \; e^{ i \sigma \varphi} = { r T_{\sigma} r' \over
r_{{}_\bot} r'_{{}_\bot} }  \, .
\label{e.10}
\end {equation}
Thus, variations of $r$ lead to rotation by angle $\varphi$ in the
2-plane $(s_1, s_2)$, and in the spin relations an additional
phase factor appears. Having fixed $r$, we thus fix the tetrad in
the Minkowski space and thus can write in explicit form the whole
set of spin operators%
\footnote[1]{ $\gamma_5 = i \gamma^0 \gamma^1 \gamma^2 \gamma^3$
can be given in form $\gamma_5 = i {\hat s}_1 {\hat s}_2 {\hat
s}_3 {\hat s}_0$.}
:
\begin {equation}
\displaystyle
\Sigma_3 = {\gamma}_5 {\hat s}_3 \, , \; \; \; \Sigma_{\sigma} =
{1 \over 2} ( {\hat s}_1 +  i \sigma {\hat s}_2 ) \, ,
\label{e.11}
\end {equation}
which act on the bispinor%
\footnote[2]{ We will keep symbols $s$ and especially $r$ only
where necessary.}
$u^{\sigma} (p, s, r)$ in following way:
\begin {equation}
\begin{array}{l}
\displaystyle
\Sigma_3 u^{\sigma} (p,s,r) = \sigma u^{\sigma} (p,s,r) \, ;
          \\       \displaystyle
\Sigma_{-\sigma} u^{\sigma} (p,s,r) = u^{-\sigma} (p,s,r) \, , \;
\;  \Sigma_{\sigma} u^{\sigma} (p,s,r) = 0 \, .
\end{array}
\label{e.12}
\end {equation}
The second relation represents the known phase agreement: the
action of the spin flip operator does not lead to the appearance
of an additional phase factor. Note the following important fact.
Having fixed the tetrad, we obtain the possibility of writing for
the 4-component bispinor four equations, namely, two equations for
the spin flip operator are added to the Dirac equation and the
equation for the spin projection.

Let us consider of the consequences of this situation. We give the
tensor $g$ in the form
\begin {equation}
\displaystyle
g^{\mu \nu} = v^{\mu} v^{\nu} - s^{\mu} s^{\nu} - s^{\mu}_1
s^{\nu}_1 - s^{\mu}_2 s^{\nu}_2  \, .
\label{e.13}
\end {equation}
Then from (\ref{e.11}), (\ref{e.13})  follows
\begin {equation}
\displaystyle
\begin{array}{l}
{\gamma}^{\mu} = g^{\mu \nu} {\gamma}_{\nu}= v^{\mu} {\hat v} -
s^{\mu} {\hat s} - s^{\mu}_1 {\hat s}_1 - s^{\mu}_2 {\hat s}_2
        \\
 = v^{\mu} {\hat v} - {\gamma}_5 \left\{ s^{\mu} \Sigma_3
 + (s^{\mu}_1 - i \sigma s^{\mu}_2) \Sigma_{\sigma}
+ (s^{\mu}_1 + i \sigma s^{\mu}_2) \Sigma_{-\sigma}
 \right\} \, .
\end{array}
\label{e.14}
\end {equation}
Using the Dirac equation and (\ref{e.12}), we have
\begin {equation}
\displaystyle
{\gamma}^{\mu} u^{\sigma} (p,s) =(  v^{\mu} - \sigma s^{\mu}
{\gamma}_5 )  u^{\sigma} (p,s) - (s^{\mu}_1 + i \sigma s^{\mu}_2)
{\gamma}_5 u^{-\sigma} (p,s)   \, .
\label{e.15}
\end {equation}
Repeated action of the $\gamma$-matrix on (\ref{e.15}) gives
\begin {equation}
\displaystyle
\begin{array}{l}
{\sigma}^{\mu \nu} u^{\sigma} (p,s) = \sigma \left( [ v \cdot
s]^{\mu \nu} {\gamma}_5 + i {\widetilde {[ v \cdot s]}}^{\mu \nu}
\right) u^{\sigma} (p,s)
        \\
- \left[ (s_1 + i \sigma s_2) \cdot ( \sigma s + {\gamma}_5 v )
\right]^{\mu \nu} u^{-\sigma} (p,s)
  \, ,
\end{array}
\label{e.16}
\end {equation}
where the notation for tensor is introduced
\begin {equation}
\displaystyle
[ a \cdot b ]^{\mu \nu} =  a^{\mu} b^{\nu} -  b^{\mu} a^{\nu} \, .
\label{e.17}
\end {equation}
Any operator $Q$ can be expanded by complete set of Dirac matrices
\begin {equation}
\displaystyle
\Gamma = \left\{ 1 , {\gamma}_5 , {\gamma}^{\mu} ,  {\gamma}_5
{\gamma}^{\mu} ,  {\sigma}^{\mu \nu} \right\}  \, .
\label{e.18}
\end {equation}
Thus, formulas (\ref{e.15}), (\ref{e.16}) provide the possibility
of reducing the action of an arbitrary matrix operator on the
bispinor: $Q u^{\sigma} (p, s)$ to expansion by a complete set of
bispinors $u^{\pm \sigma} (p, s)$, $\gamma_5 u^{\pm \sigma} (p,
s)$. The expansion coefficients will be tensors whose rank is
equal to the number of free Lorentz-indices in the operator $Q$.

%%%%%%%%%%%%%%%%%%%%%%%%%%%%%%%%%%%%%%%%%%%%%%%%%%%%%%%%%%%%%%%%%%%%%%%%

\section {The transit matrices between the different spin bases}%3

Let us assume that we know the set of amplitudes in a certain spin
basis
\begin {equation}
\displaystyle
{\cal M}_{\delta' \delta} =  \bar{u}^{{\delta}'} ( p' , s' ) Q
u^{\delta}( p , s ) \, ,
\label{e.19}
\end {equation}
but the initial state is represented so that the axis of the spin
projections is the vector $l$. Thus, we are interested in the
amplitudes
\begin {equation}
\displaystyle
{\cal M}'_{\delta' \sigma} =  \bar{u}^{{\delta}'} ( p' , s' ) Q
u^{\sigma}( p , l ) \, ,
\label{e.20}
\end {equation}
or to be more exact, in the final analysis, the squares of their
modules.

Using the equation
\begin {equation}
\displaystyle
\sum\limits_{\delta}^{} {1 \over 4} (1 + \hat{v}) (1 + {\delta}
{\gamma}_5 \hat{s} ) u^{\sigma} ( p , l ) = u^{\sigma} ( p , l )
\label{e.21}
\end {equation}
we obtain the expression for the squared module of the amplitude
(\ref{e.20}) (arguments in bispinors are omitted):
\begin{eqnarray}
\displaystyle
\begin{array}{c}
\displaystyle |{\cal M}'_{{\delta}' {\sigma}}|^2 =
\bar{u}^{{\delta}'} Q u^{\sigma} \; \bar{u}^{\sigma} \bar{Q}
u^{{\delta}'}
\\  \displaystyle
= \sum\limits_{{\delta}_1 {\delta}_2}^{} \bar{u}^{{\delta}'}  Q {1
\over 4} (1 + \hat{v}) (1 + {\delta}_1 {\gamma}_5 \hat{s})
u^{\sigma} \bar{u}^{\sigma}{1 \over 4} (1 + \hat{v}) (1 +
{\delta}_2 {\gamma}_5 \hat{s}) \bar{Q} u^{{\delta}'}
\\ \displaystyle
= \sum\limits_{{\delta}_1 {\delta}_2}^{} \bar{u}^{{\delta}'} Q
u^{{\delta}_1}  \;  \bar{u}^{{\delta}_1} u^{\sigma}
\bar{u}^{\sigma} u^{{\delta}_2} \; \bar{u}^{{\delta}_2} \bar{Q}
u^{{\delta}'} = \sum\limits_{{\delta}_1 {\delta}_2}^{} {\cal
M}_{{\delta}' {\delta}_1} {\cal K}_{{\delta}_1 {\delta}_2} {\cal
M}^{*}_{{\delta}' {\delta}_2} \; ,
\end{array}
\label{e.22}
\end{eqnarray}
where
\begin {equation}
\displaystyle
{\cal K}_{{\delta}_1 {\delta}_2} = \bar{u}^{{\delta}_1} u^{\sigma}
\bar{u}^{\sigma} u^{{\delta}_2} = \Tr u^{\sigma} \bar{u}^{\sigma}
u^{{\delta}_2}  \bar{u}^{{\delta}_1} \ .
\label{e.23}
\end {equation}
Thus, if we know the explicit form of matrix (\ref{e.23}), then
using (\ref{e.19}), we calculate the squared modules of amplitudes
(\ref{e.20}).

If $\delta_1 = \delta_2 = \delta$, then \cite{r.10}
\begin {equation}
\displaystyle
{\cal K}_{{\delta} {\delta}} = {1 \over 16} \Tr  (1 + \hat{v}) (1
+ {\sigma} {\gamma}_5 \hat{l} ) (1 + \hat{v}) (1 + {\delta}
{\gamma}_5 \hat{s} ) = {1 \over 2} \left( 1 - {\sigma} {\delta} (l
s) \right) \ .
\label{e.24}
\end {equation}

Using the operators of spin flipping \cite{r.10}
$\displaystyle {1 \over 2} {\gamma}_5 ( {\hat{s}}_1 - i {\delta}
{\hat{s}}_2 )$,
we obtain the expression for the configuration
$\displaystyle {\delta}_1 = - {\delta}_2 = \delta$:
\begin{eqnarray}
\displaystyle
\begin{array}{l}
\displaystyle {\cal K}_{{\delta} -{\delta}} = {1 \over 32} \Tr  (1
+ \hat{v}) (1 + {\sigma} {\gamma}_5 \hat{l} ) {\gamma}_5 (
{\hat{s}}_1 - i {\delta} {\hat{s}}_2 ) (1 + \hat{v}) (1 + {\delta}
{\gamma}_5 \hat{s} )
\\
\displaystyle = {1 \over 16} \Tr {\gamma}_5 ( {\hat{s}}_1 - i
{\delta} {\hat{s}}_2 ) ( {\sigma} {\gamma}_5 \hat{l}  + {\sigma}
{\delta} \hat{s} \hat{v} \hat{l} ) = - {\sigma \over 2} ( s_1 - i
{\delta} s_2 ) l \ .
\end{array}
\label{e.25}
\end{eqnarray}
In deriving (\ref{e.25}), we used the representation $(s_3 =s, s_0
= v)$
\begin {equation}
\displaystyle
\gamma_5 = i {\hat s}_1 {\hat s}_2 {\hat s}_3 {\hat s}_0 \, .
\label{e.26}
\end {equation}
>From the explicit form (\ref{e.24}), (\ref{e.25}) of the elements
of the matrix ${\cal K}_{\delta_1 \delta_2}$ it follows that it
can be written with the aid of Pauli matrices $\sigma_i$:
\begin {equation}
\displaystyle
{\cal K}_{\delta_1 \delta_2} = \left( 1 - \sigma \sigma_i (l s_i)
\right)_{\delta_1 \delta_2} \, .
\label{e.27}
\end {equation}
(\ref{e.27}) can be given in a covariant 4-dimensional form. To do
this, we introduce the isotropic 4-vector
\begin {equation}
\displaystyle
L = \left( 1 ;  \sigma (l s_i) \right)
\label{e.28}
\end {equation}
and take into account that $\sigma_0 = {\bf 1}$. Than
\begin {equation}
\displaystyle
{\cal K} = \sigma_{\mu} L^{\mu} \; .
\label{e.29}
\end {equation}
It should be noted that $1$ is nothing but the square of the
4-velocity of the particle, and $l$ and $s_3$ are the axes of spin
projections in different spin bases.

Note that into (\ref{e.27}), (\ref{e.28}) enter only the vectors
of the tetrad in which we calculated amplitudes (\ref{e.19}). It
can be shown that the finite expression (\ref{e.22}) is
independent of not only the phase vector $r$, but also the vectors
$s_1$ and $s_2$, since they enter into (\ref{e.22}) in
combinations
$s_1^{\mu} s_1^{\nu} + s_2^{\mu} s_2^{\nu} = s_3^{\mu} s_3^{\nu} -
g ^{\mu \nu}$
and
 $\varepsilon^{\mu \nu \rho \sigma} {s_1}_{\rho}
{s_2}_{\sigma} = [ v_0 \cdot s_3 ]^{\mu \nu}$.

Using formulas (\ref{e.24}), (\ref{e.25}), (\ref{e.27}),
(\ref{e.29}) we can construct a matrix $\cal K$ for any fermion
participating in the reaction. And the representation of
(\ref{e.27}), (\ref{e.29}) provides the possibility of using, if
necessary, the Fierz transformation.

Note the following important fact. Formula (\ref{e.22}) can also
be used in the case where the beam is partially polarized and the
states are not pure. To do this, it is enough to carry out
substitution $\sigma l \to a$, $|a^2|<1$ in (\ref{e.24}),
(\ref{e.25}), (\ref{e.28}). $a$ is the vector of partial
polarization (see \cite{r.6}).

To investigate the polarization phenomena, besides the squared
modules of amplitudes (\ref{e.22}), it is necessary to know the
values of $  {\cal M}_{\sigma} {\cal M}^{*}_{- {\sigma}}  $,%
\footnote[3]{The projection of only that particle whose
polarization we are investigating is given in the amplitude
designations.}
since various polarization characteristics are expressed in their
real and imaginary parts.

Thus, in the general case, it is necessary to know the matrix
\begin {equation}
\displaystyle
{\cal K}^{ {\sigma}_1 {\sigma}_2 }_{ {\delta}_1 {\delta}_2 } = \Tr
u^{ {\sigma}_1 } \bar{u}^{{\sigma}_2} u^{ {\delta}_2 }
\bar{u}^{{\delta}_1} \, ,
\label{e.30}
\end {equation}
and at $\sigma_2 =\sigma_1$, we obtain (\ref{e.22}), (\ref{e.23}).

It can be shown that the matrix $\cal K$ is of the form
($\delta_{\alpha \beta}$ are Kroneker symbols)
\begin {equation}
\displaystyle
{\cal K}^{ {\sigma}_1 {\sigma}_2 }_{ {\delta}_1 {\delta}_2 } = {1
\over 2} \left\{ \delta_{{\sigma}_1 {\sigma}_2} \delta_{{\delta}_1
{\delta}_2} - \sigma_i^{ {\sigma}_2 {\sigma}_1 } \sigma_j^{
{\delta}_1 {\delta}_2 } (l_i s_j)  \right\} \, .
\label{e.31}
\end {equation}
It should be recalled that with respect to $\delta_1$ and
$\delta_2$ summation with the amplitudes calculated in the spin
basis with the tetrade $\{v , s_3 , s_1 , s_2\}$ is made.

   In conclusion, note that the formula
\begin {equation}
\displaystyle
{{\cal M}'}_{{\sigma}_1} {{\cal M}'}_{{\sigma}_2}^{*} =
\sum\limits_{ {\delta}_1 \, {\delta}_2}^{} {\cal M}_{{\delta}_1}
{\cal K}^{ {\sigma}_1 {\sigma}_2 }_{ {\delta}_1 {\delta}_2 } {\cal
M}_{{\delta}_2}^{*}
\label{e.32}
\end {equation}
represent a new way of describing the transition from one spin
basis to another without using the extremely cumbersome and
noncovariant apparatus of $D$-Wigner functions.

%%%%%%%%%%%%%%%%%%%%%%%%%%%%%%%%%%%%%%%%%%%%%%%%%%%%%%%%%%%%%%%%%%%%%%%%

\section {Diagonal spin basis (DSB)}%4

In any process, only an even number of fermions can participate.
If the number of free fermions is $2n$, each diagram contains $n$
open fermion lines and is described by a structure consisting of
contractings between $n$ fermion ``sandwiches'' of the type
\begin {equation}
\displaystyle
\bar{\Phi}^{{\sigma}'} ( p' , s' ) Q {\Phi}^{\sigma}( p , s ) \, .
\label{e.33}
\end {equation}
For the fermion line $\Phi$ are bispinors $u$ (see (\ref{e.1}))
and for the antifermion line -- bispinors $v$; for annigilation
and for creation of an pair $\Phi$ are bispinors $u$ and $v$.

For certainty, we shall consider the structures (\ref{e.1}),
(\ref{e.2}).

We shall not consider the group aspects of introduction of the
DSB. We only note that in DSB the vectors $s$ and $s'$ are chosen
such that they belong to the 2-plane $(p, p')$ or $(v, v')$;
$\displaystyle v = {p \over m}$, $\displaystyle v' = {p' \over
m'}$. Satisfying this requirement, we obtain
\begin{equation}
\displaystyle
s = { (v v') v - v' \over \sqrt{ (v v')^2 - 1 }} \; , \; \;
s' = - { (v v') v' - v \over \sqrt{ (v v')^2 - 1 }} \; .
\label{e.34}
\end{equation}

Thus, the support vectors (see (\ref{e.3})) for the {\em on-} and
{\em off-} states of the fermion line are vectors $q=v'$, $q'=-v$.
With such a choice of signs in special reference systems vectors
$\vec s$ and $\vec {s'}$ coincide with the direction of the
3-momentum of the initial particle and  are opposite to the
3-momentum of the final particle. By special reference systems is
meant the Breit system for fermion or antifermion lines
($t$-lines) and s.c.m. for the pair being annihilated or created
($s$-lines).

>From (\ref{e.34}) it can easily be seen that in the derived
reference systems diagonality gets the meaning of helicity, with
$\delta =\lambda $, ${\delta}' = - {\lambda}' $. Thus, the DSB is
covariant description of the helicity in the special reference
systems. Exactly in these systems helicity the pairs of particles
as well as such notions as non-flip and flip amplitudes have a
clear physical meaning.  Indeed, in helicity basis in arbitrary
reference system the spin of the initial particle is projected on
the 3-momentum $\vec p$ and that of the final particle on $\vec
{p'} $, than what can be said about the non-flip or flip process?
These notions in helicity basis have a only marking meaning.

In our opinion, neglect of this fact is the main reason why the
process of constructing operators (\ref{e.2}) convenient for
calculation has been extended to decades. Attempts to construct
covariant operators (\ref{e.2}) in helicity basis look
unreasonable. The helicity of massive particle is a ``bad''
quantum number, since it is not invariant at a Lorentz
transformation. Any declaration of covariancy of operators
(\ref{e.2}) in the helicity basis usually is a disguised
transition to the special reference system. It should be noted
that other fermion pair ``droop'', since each pair has its own
special reference system. The only exception is $e^{+} e^{-} \to
\mu^{+} \mu^{-}$ type reactions. No wonder that just these
reactions are chosen, as a rule, for examples of calculations of
processes.

Before to go to the construction of tetrads for the initial and
final states, we introduce into the 2-plane $(v,v')$ two $v$- and
$v'$- symmetrized, orthonormalized vectors
\begin{equation}
\displaystyle
n_0 = {v + v' \over 2 V_{+} } \ , \;
n_3 = {v - v' \over 2 V_{-} } \ ;
\label{e.35}
\end{equation}
\begin{equation}
\displaystyle
V_{\pm} = \sqrt{v v' \pm 1 \over 2  } \; .
\label{e.36}
\end{equation}

>From (\ref{e.5}), (\ref{e.34}), (\ref{e.35}) it follows that
\begin {equation}
\begin{array}{l}
\displaystyle
g^{\mu \nu}_{\parallel} = v^{\mu} v^{\nu} - s^{\mu} s^{\nu}  =
{v'}^{\mu} {v'}^{\nu} - {s'}^{\mu} {s'}^{\nu} = n_0^{\mu}
n_0^{\nu} - n_3^{\mu} n_3^{\nu} \, ,
       \\  \displaystyle
 \tilde{\varepsilon}^{\mu \nu}_{\parallel} =
{\varepsilon}^{\mu \nu \rho \sigma} v_{\rho} s_{\sigma}
={\varepsilon}^{\mu \nu \rho \sigma} {v'}_{\rho} {s'}_{\sigma} =
{\varepsilon}^{\mu \nu \rho \sigma} {n_0}_{\rho} {n_3}_{\sigma} =
{1 \over 2 V_{+} V_{-} } {\varepsilon}^{\mu \nu \rho \sigma}
v_{\rho} {v'}_{\sigma} \, .
\end{array}
\label{e.37}
\end {equation}

>From (\ref{e.6}), (\ref{e.37}) it follows that the choice of
common phase vector $r'=r$ leads to the coincidence for both
vectors of the tetrads lying in the orthogonal 2-plane, i.e.
\begin {equation}
\displaystyle
n_1^{\mu} = s_1^{\mu} = {s'}_1^{\mu} =( g^{\mu \nu}_{\parallel} -
g^{\mu \nu}) {r_{\nu} \over r_{{}_\bot} } \, ,  \; \; \; n_2^{\mu}
= s_2^{\mu} = {s'}_2^{\mu} = -\tilde{\varepsilon}^{\mu
\nu}_{\parallel} {r_{\nu} \over r_{{}_\bot} } \, .
\label{e.38}
\end {equation}

Next, let us coincidence the plane Lorentz transformation that
transforms $v$ to $v'$. In the representation of the group
$SL(2,C)$, it is of the form
\begin {equation}
\displaystyle
\Lambda (v \to v') = { 1 + \hat{v'} \hat{v} \over 2 V_{+} }
 \, ,
\label{e.39}
\end {equation}
and $\Lambda \hat v  \Lambda^+ =\hat{v'} $. From (\ref{e.34}) it
also follows that $\Lambda \hat{s}  \Lambda^{+} = \hat{s'} $.
Transformation (\ref{e.39}) does not change vectors lying in the
orthogonal 2-plane. Thus, the Lorentz transformation (\ref{e.39})
converts the tetrad of the initial particle into the tetrad of
final particle:
\begin {equation}
\displaystyle
\Lambda (v \to v') \left\{ \hat{v}, \hat{s}, \hat{n}, \hat{n}_2
\right\} \Lambda^{+} (v \to v') = \left\{ \hat{v'}, \hat{s'},
\hat{n'}, \hat{n'}_2 \right\}
 \, ,
\label{e.40}
\end {equation}
and this in turn means that the relation between the bispinors of
the initial and final states in DSB is of the form
\begin {equation}
\displaystyle
u^{\delta} (p', s') = \Lambda (v \to v') u^{\delta} (p, s) \, .
\label{e.41}
\end {equation}

In the DSB we choose the normalization
\begin {equation}
\displaystyle
\bar{u}^{\delta} (p) u^{\delta} (p) =  \bar{u}^{{\delta}'} (p')
u^{{\delta}'} (p') = 1 \, .
\label{e.42}
\end {equation}
Then the relation (\ref{e.41}) describes the cases $m' \not= m$
too.

To restore the generally accepted normalization, it is necessary
to multiply the amplitudes calculated in the DSB by factor
\begin {equation}
\displaystyle
\prod\limits_{ i= 1}^{n} \sqrt{ 2 m_i 2 m'_i} \, ,
\label{e.43}
\end {equation}
where $n$ is the number of open fermion lines.

Using the Dirac equation and (\ref{e.14}), we can write relation
(\ref{e.41}) in different representations
\begin {equation}
\displaystyle
u^{\delta} (p' , s') = { \hat{v'} + 1 \over 2 V_{+} } u^{\delta}
(p, s ) =  {\hat n}_0 u^{\delta} (p, s ) = ( V_{+} - \delta
{\gamma}_5 V_{-} ) u^{\delta} (p, s ) \, .
\label{e.44}
\end {equation}

Formula (\ref{e.44}) makes it possible to express in the DSB the
explicit form of the transition operator (\ref{e.2}) in terms of
the projective operators of the initial (or final) state
\begin {equation}
\displaystyle
 u^{\delta} (p, s ) \bar{u}^{\delta} (p, s) = {1 \over 4} ( {\hat
 v} + 1) ( 1 + \delta {\gamma}_5 {\hat s}) \, .
\label{e.45}
\end {equation}

As relation (\ref{e.44}), the transition operators (\ref{e.2}) can
be given in different form. Some of them are shown below.
\begin {equation}
\begin{array}{l}
\displaystyle
4 u^{\delta} (p, s) \bar{u}^{\delta} (p', s') = ( {\hat v} + 1)
\left( {1 \over 2 V_{+} } - {\delta {\gamma}_5 \over 2 V_{-} }
\right) ( {\hat v}' + 1) =
            \\  \displaystyle
= ( V_{+} + \delta {\gamma}_5 V_{-} ) ( 1 - \delta {\gamma}_5
{\hat n}_0 {\hat n}_3 ) + {\hat n}_0  + \delta {\gamma}_5 {\hat
n}_3 =
           \\  \displaystyle
= \left( 1 + {1 \over 2} ( V_{+} + \delta {\gamma}_5 V_{-} )
({\hat n}_0  + \delta {\gamma}_5 {\hat n}_3) \right) ({\hat n}_0
+ \delta {\gamma}_5 {\hat n}_3)  \, ,
\end{array}
\label{e.46}
\end {equation}
\begin {equation}
\begin{array}{c}
\displaystyle
4 u^{\delta} (p, s) \bar{u}^{- \delta} (p', s') =
          \\  \displaystyle
= {\delta \over r_{\perp} } ( {\hat v} + 1) \left( {1 \over 2
V_{-} } ( {\hat r} - { r ( v + v') \over v v' + 1 } )
 - {\delta {\gamma}_5 \over 2 V_{+} }
( {\hat r} - { r ( v - v') \over v v' - 1 } )
 \right) ( {\hat v}' + 1) =
            \\  \displaystyle
= {\gamma}_5 ( V_{+} + \delta {\gamma}_5  V_{-} - {\hat n}_0 )
({\hat n}_1 + i \delta  {\hat n}_2)=
           \\  \displaystyle
= {\gamma}_5 \left( V_{+} + \delta {\gamma}_5  V_{-} - {1 \over 2}
({\hat n}_0  + \delta {\gamma}_5 {\hat n}_3) \right) ({\hat n}_1 +
i \delta  {\hat n}_2)  \, .
\end{array}
\label{e.47}
\end {equation}

In deriving formula (\ref{e.47}) from (\ref{e.46}), the spin flip
operator (\ref{e.11}) is used and the relation valid for an
arbitrary orthonormalized tetrad also hold:
\begin {equation}
\displaystyle
{\hat n}_0 ({\hat n}_1 + i \delta  {\hat n}_2) = \delta {\gamma}_5
{\hat n}_3 ({\hat n}_1 + i \delta  {\hat n}_2) \, , \; ({\hat n}_0
+ \delta {\gamma}_5 {\hat n}_3)^2 = 2 ( 1 - \delta {\gamma}_5
{\hat n}_0 {\hat n}_3 ) \, .
\label{e.48}
\end {equation}

Since any interaction operator can be expanded by the complete set
of Dirac matrices (\ref{e.18}), it is tempting to calculate the
matrix elements of this set in DSB. From (\ref{e.1}),
(\ref{e.46}), (\ref{e.47}) it follows that
\begin {equation}
\begin{array}{c}
\displaystyle
\bar{u}^{\delta} (p', s')  \left\{ 1 ; {\gamma}_5 ; {\gamma}^{\mu}
;  {\gamma}_5 {\gamma}^{\mu} ;  {\sigma}^{\mu \nu} \right\}
u^{\delta} (p, s) =
          \\  \displaystyle
= \left\{ V_{+} \; ; \; \delta V_{-} \; ; \; n_0^{\mu} \; ; \; -
\delta n_3^{\mu} \; ; \; V_{-} [ n_0 \cdot n_3]^{\mu \nu} - i
\delta V_{+} {\widetilde {[ n_0 \cdot n_3]}}^{\mu \nu} \right\} \,
,
\end{array}
\label{e.49}
\end {equation}
\begin {equation}
\begin{array}{c}
\displaystyle
\bar{u}^{- \delta} (p', s')  \left\{ 1 ; {\gamma}_5 ;
{\gamma}^{\mu} ;  {\gamma}_5 {\gamma}^{\mu} ;  {\sigma}^{\mu \nu}
\right\} u^{\delta} (p, s) =
          \\  \displaystyle
= \left\{ 0 \; ; \; 0 \; ; \; \delta V_{-} ({\hat n}_1 + i \delta
{\hat n}_2)^{\mu} \; ; \; - V_{+} ({\hat n}_1 + i \delta {\hat
n}_2)^{\mu} \; ; \;  \delta \left[ n_3 \cdot ({\hat n}_1 + i
\delta  {\hat n}_2) \right]^{\mu \nu} \right\} \, .
\end{array}
\label{e.50}
\end {equation}

The matrix elements (\ref{e.49}), (\ref{e.50}) can be interpreted
as spin characteristics of exchange particles under the scalar,
pseudoscalar, axial and tensor interaction, respectively.

>From (\ref{e.50}) it follows that the exchange (emission,
absorption) of a scalar or pseudoscalar particle does not change
the fermion spin projection in DSB. Thus, the choice of DSB not
only leads to simple calculation formulas, but also adequately
reflects the physical meaning of the processes being described. To
support this thesis, we consider in DSB the matrix elements of
nucleon current
\begin {equation}
\displaystyle
J^{\delta ,\delta'}_\mu ={\bar u}^{\delta'} (p',s') \left( F_1
(q^2) \gamma_{\mu} - {F_2 (q^2) \over 2m} \sigma_{\mu \nu} q^{\nu}
\right) u^{\delta} (p,s) \, ; \;  q = p' - p \, .
\label{e.51}
\end {equation}
>From (\ref{e.49}), (\ref{e.50}) we have
\begin {equation}
\displaystyle
J^{\delta ,\delta}_\mu = G_E {n_0}_{\mu} \, , \; J^{\delta
,-\delta}_\mu = \delta V_{-} G_M ( n_1 + i \delta n_2 )_{\mu} \, .
\label{e.52}
\end {equation}

It is noteworthy that in DSB formfactors $F_1$ and $F_2$
independently go in Saks formfactors having a clear physical
meaning:
\begin {equation}
\displaystyle
G_E = F_1 + { q^2 \over 4 m^2} F_2 \, , \; G_M = F_1 + F_2 \, ,
\label{e.53}
\end {equation}
$G_E$ and $G_M$ describe the distribution of the electric charge
and the magnetic moment of nucleon, respectively. Thus, from
(\ref{e.52}) it follows that the situation where the non-flip
processes are responsible for the electric interaction and the
flip spin processes -- for the magnetic interaction is covariantly
described in DSB. It should be noted that the space-like ($q^2
<0$) virtual photon has a scalar or a circular polarization,
respectively.

In helicity basis, grouping of (\ref{e.53}) occurs only in Breit's
system in which the Fourier transformations becomes
three-dimensional since
$$\exp (- i q r) = \exp i ({\vec q} {\vec r} - q_0 t) = \exp (i
{\vec q} {\vec r}) \, .$$

Formulas (\ref{e.46}), (\ref{e.47}) described the fermion
$t$-line. To describe the antifermion line as well as the $s$-line
for the annihilating and creating pairs, one should make use by
the relation
\begin {equation}
\displaystyle
v^{\delta} (p,s) = - \delta {\gamma}_5 u^{- \delta} (p,s) \, ,
\label{e.54}
\end {equation}
which relates the particle and antiparticle bispinors.

If in (\ref{e.46}), (\ref{e.47}) we restore by the recipe
(\ref{e.43}) the normalization
$\bar{u}^{\delta} (p) u^\delta (p) = 2m$
and perform the limiting transition $m \to 0$ or/and $m' \to 0$,
we will obtain the transition operators for processes involving
massless fermions found in \cite{r.7}.

For calculating concrete processes, it may be convenient to
utilize a formalism, in which the basis spinor
$u^\delta (n_0 , n_3 ; n_1 , n_2)$
common for initial and final bispinors and satisfying the
conditions
\begin {equation}
\displaystyle
{\hat n}_0 u^{\delta} (n_0 , n_3) =  u^{\delta} (n_0 , n_3) \, ,
\; {\gamma}_5 {\hat n}_3  u^{\delta} (n_0 , n_3) = \delta
u^{\delta} (n_0 , n_3) \ ,
\label{e.55}
\end {equation}
is used.

It is easy to see%
\footnote[4]{For this the relation $\displaystyle { V_{-} \over
V_{+} + 1} = { V_{+} - 1 \over V_{-} } $ is used.}
that plane Lorentz transformations
$\displaystyle \Lambda ( n_0 \to v) = { 1 + {\hat v} {\hat n}_0
\over \sqrt{ 2 (V_{+} + 1 )}} $
and
$\displaystyle \Lambda ( n_0 \to v') = { 1 + {\hat v}' {\hat n}_0
\over \sqrt{ 2 (V_{+} + 1 )}}  $
change the basis spinor tetrad to tetrads of the initial and final
states respectively. Therefore, the transition operators
(\ref{e.46}), (\ref{e.47}) can be given in the form
\begin {equation}
\begin{array}{c}
\displaystyle
4 u^{\delta} (p, s) \bar{u}^{\delta'} (p', s') =  {2 \over V_{+}
+1} ( {\hat v} + 1) u^{\delta} (n_0 , n_3) {\bar u}^{\delta'} (n_0
, n_3) ( {\hat v}' + 1) =
        \\    \displaystyle
= {1 \over 2(V_{+} +1)} ( {\hat v} + 1)( {\hat n}_0 + 1)
({\delta}_{\delta' \delta} + {\gamma}_5 {\hat n}^i
{\sigma}_{\delta' \delta}^i  ) ( {\hat v}' + 1) \, .
\end{array}
\label{e.56}
\end {equation}
In this equality the Bouchiat and Michel relation \cite{r.8} is
used.

In (\ref{e.56}), vectors $v$ and $v'$ can be expanded in vectors
$n_0$ and $n_3$ with the aid of relations (\ref{e.35}).

In conclusion of this section, we give the recipe for calculating
in DSB exchange diagrams. As an example, we consider the
electron-electron scattering. For certainty, let particles $1,\,
3$ and $2,\, 4$ are paired. Then the exchange diagram has the
structure
\begin {equation}
\displaystyle
{\bar u}^{{\delta}_4} (p_4) {\gamma}_{\mu} u^{{\delta}_1} (p_1)
{\bar u}^{{\delta}_3} (p_3) {\gamma}^{\mu} u^{{\delta}_2} (p_2) =
\Tr {\gamma}_{\mu} u^{{\delta}_1} (p_1) {\bar u}^{{\delta}_3}
(p_3) {\gamma}^{\mu}  u^{{\delta}_2} (p_2) {\bar u}^{{\delta}_4}
(p_4) \, ,
\label{e.57}
\end {equation}
i.e. it is expressed in terms of the transition operators entering
into direct diagram.

%%%%%%%%%%%%%%%%%%%%%%%%%%%%%%%%%%%%%%%%%%%%%%%%%%%%%%%%%%%%%%%%%%%%%%%%

\section{Examples of calculations in DSB}%5

In Section {\bf 4}, it is shown how the main task of the approach
based on the use of formula (\ref{e.1}) is solved in DSB in
compact, covariant form. Namely, the problem of finding in
explicit form the transition operators (\ref{e.2}). In each
particular reaction, a further simplification of calculations can
be achieved by simplifying the form of the interaction operators
$Q$. Let us demonstrate this by an example of calculating the
processes of radiation of bremsstrahlung photons. Let the initial
fermion in the state $u^\delta (p,s,r)$ emits a photon with a
momentum $k$ and helicity $\lambda$. We do not concentrize the
spin projection vector of the fermion so far.

Choose the photon polarization vector in the form \cite{r.9}
\begin {equation}
\displaystyle
e^\lambda (k) = {1 \over {\sqrt{2} \, k_\perp}}
(\varepsilon_\parallel +  i \lambda {\tilde
\varepsilon}_{\parallel} - \lambda \sigma) k \, ,
\label{e.58}
\end {equation}
\begin {equation}
\displaystyle
{\varepsilon_{\parallel} \ }^{\mu \nu} = v^{\mu} s^{\nu} - s^{\mu}
v^{\nu} \, , \quad {{\tilde \varepsilon}_{\parallel} \ }^{\mu \nu}
= {1 \over 2} \varepsilon^{\mu \nu \rho \sigma}
{\varepsilon_{\parallel}}_{\mu \nu} \, .
\label{e.59}
\end {equation}

It is easy to see that vector (\ref{e.58}) meets the necessary
requirement for the polarization vector:
$ke^\lambda =(e^\lambda)^2=0$, $e^\lambda e^{-\lambda}=- 1$.
>From the last equation in (\ref{e.12}) and condition (\ref{e.10})
it follows
\begin {equation}
\displaystyle
 i \gamma {\tilde \varepsilon}_{\parallel} k u^\sigma (p,s) = -\sigma
 \gamma (g_\parallel - g) k u^\sigma (p,s) \, ,
\label{e.60}
\end {equation}
therefore, taking into account that
$\displaystyle e^{\lambda} v = { (s- \lambda \sigma v) k \over
{\sqrt{2} \, k_\perp}} $,
we obtain
\begin {equation}
\displaystyle
    {\hat e}^{\lambda} u^\sigma (p,s) = e^{\lambda}
    v (1 + \lambda \gamma_{5}) u^\sigma (p,s) \, .
\label{e.61}
\end {equation}

In amplitude, the operator ${\hat e}^{\lambda} $ is followed by
operator $\hat p -\hat k +m$. Using (\ref{e.61}) and the
commutation relation
${\hat p} {\hat e} = 2 p e - {\hat e} {\hat p} $,
it can easily be seen that
\begin {equation}
\displaystyle
  (\hat{p} - \hat{k} + m) \hat{e}^\lambda u^\sigma (p,s)
  = 2 e^\lambda  p \left( 1 - {1 \over 2m} (1 -
  \lambda \gamma_{5}) \hat{k} \right) u^\sigma (p,s) \, .
\label{e.62}
\end {equation}

The fact is rather surprising that if $n$ photons with equal
helicity $\lambda$ are emitted from the fermion line, then the
structure avalanche-like increasing in complexity acquires in this
approach a very trivial form
\begin {equation}
\begin{array}{c}
\displaystyle
(\hat{f_n} + m) \hat{e_n} \cdots (\hat{f_1} + m) \hat{e_1} \,
u^\sigma (p,s) =
        \\  \displaystyle
=  2^n f_n e_n \cdots f_1 e_1 \left( 1 - {1 \over 2m} (1 - \lambda
\gamma_{5}) \hat{k} \right) u^\sigma (p,s) \, ,
\end{array}
\label{e.63}
\end {equation}
here $f_i = p - k_i - \ldots - k_1 $ is the momentum of  fermion
between the photons $k_i$ and $k_{i+1}$, $k = k_1 + \ldots + k_n
$.

We now consider the following problem. In those diagrams where a
photon is emitted from the final fermion line, we shall still use
vector (\ref{e.58}). However, we shall not obtain a structure
similar to (\ref{e.62}) because tensors (\ref{e.59}) were
constructed for the initial fermion line. This problem disappears
in DSB where these tensors coincide (\ref{e.37}).

If a photon is emitted from another fermion line, then by virtue
of gauge invariance we can make the following substitution in
these diagrams:
\begin {equation}
\displaystyle
e^{\lambda} \to e^{-i \lambda \varphi} e^{\prime \lambda}  \, , \;
e^{i \lambda \varphi} =  e^{* \lambda}    e^{\prime \lambda} \, ,
\label{e.64}
\end {equation}
(see \cite{r.9}) where the vector $e'$ constructed by the recipe
of (\ref{e.58}) on basis vectors of another fermion line.

If we pass to limit of massless fermions, we will obtain the known
results CALCUL-group \cite{r.11}, including the factorization of
the sum of the contributions of all the diagrams obtained on
permutation of photons in (\ref{e.63}).

If considering the radiation of gluons, it is necessary to take
into account details associated with the gauge invariance in
non-Abel theory \cite{r.12}.

In the processes involving two bremsstrahlung photons, the case
$\lambda_1 =\lambda_2 =\lambda$ is described by vector
(\ref{e.58}). For the polarization $\lambda_1 =-\lambda_2
=\lambda$, it is possible to construct a polarization vector
common for two photons by formulas analogous to (\ref{e.38}). In
this case, the amplitude structure takes on the form of a
one-photon process and fact of the participation of two photons is
only reflected by the kinematics factor \cite{r.13}.

In \cite{r.13} and review \cite{r.1}, a number of processes going
to field of an intense laser wave are considered. One of these
processes, $e^\pm +n{\gamma}^\ast =e^\pm +\gamma$ ($n$ is number
of coherently absorbing laser photons) is basic in the operation
of $e\gamma$ and $\gamma \gamma$ colliders \cite{r.14}, the others
are calibration and test ones. Calculation at level of amplitudes
makes it possible to investigate the nonlinear and spin effects.
And the spin effects are especially significant exactly in the
nonlinear processes.

If reaction involves $W^\pm$, $Z^0$ bosons, the structure of the
transition operators  (\ref{e.46}), (\ref{e.47}) is considerably
simplified. This is due to the relations of the type
$$(1\pm \gamma_5)({\hat n}_0 + \delta \gamma_5{\hat n}_3) = (1\pm
\gamma_5)({\hat n}_0 + \delta {\hat n}_3) \, ,$$
i.e. these operators are expressed in terms isotropic tetrads, and
the relation $(1\pm \gamma_5)(1\mp \gamma_5)=0$ reduces to zero
some of the terms.

If two bosons participate in the process, their tetrads are
plotted in the same manner as for a pair of fermions. In so doing,
the circular polarization vectors of bosons coincide. With the
participation of tree bosons, as, for example, in the reaction
$e^+ e^- \to W^+ W^- Z^0$, it is convenient to choose for them a
common support vector $q = k_1 + k_2 + k_3 = p_1 + p_2$ and a
common vector $n_2$. And
${n_2}^\mu \sim \varepsilon^{\mu \nu \rho \sigma} k_{1 \nu}k_{2
\rho} k_{3 \sigma}$.
This will considerably simplify the interaction operator $Q$ whose
structure comprises up to 5 $\gamma$-matrices.

%%%%%%%%%%%%%%%%%%%%%%%%%%%%%%%%%%%%%%%%%%%%%%%%%%%%%%%%%%%%%%%%%%%%%%%%

%REFERENCES

\end{document}